\documentclass[reprint,amsmath,amssymb,aps,pra]{revtex4-1}
\usepackage[utf8]{inputenc}
\usepackage{graphicx}
\usepackage{dcolumn}
\usepackage{bm}
\def\ph2{{\it p}-H$_2$}
\def\Am2{\AA$^{-2}$}

\begin{document}
\title{Adsorption of parahydrogen on graphene}
\author{Marisa Dusseault$^\star$ and Massimo Boninsegni}
\affiliation{Department of Physics, University of Alberta, Edmonton, Alberta, T6G 2E1, Canada}
\date{\today}
\begin{abstract}
We study the low temperature properties of a single layer of parahydrogen adsorbed on graphene, by means of Quantum Monte Carlo simulations. The computed phase diagram is  very similar to that of helium on the same substrate, featuring commensurate solid phases with fillings 1/3 and 7/16, as well as domain wall phases at intermediate coverages. At  higher coverage the system transitions to an incommensurate, compressible phase. Evidence of promotion of molecules to the second layer is observed at a coverage $\sim 0.112$ \Am2, significantly above existing theoretical estimates.
\end{abstract}
\maketitle
\section{Introduction}
Adsorption of highly quantal fluids on novel forms of carbon, such as nanotubes \cite{ajp}, as well as graphene (which is a single sheet of graphite) has elicited considerable attention. The theoretically predicted phase diagram of $^4$He adsorbed on graphene closely resembles that on graphite \cite{bretz,hering,schick,carneiro,ecke,greywall,crowell,crowell2,cole,pierce,corboz}, the weaker attraction coming from a single sheet not causing any significant physical change \cite{kwon,happacher}. Specifically, as a function of coverage (2D density) the first $^4$He adlayer forms two commensurate crystalline phases, registered with the underlying substrate, at coverage 1/3 (henceforth referred to as C$_{1/3}$) and 7/16 (C$_{7/16}$), separated by a domain wall phase. A first order phase transition to an incommensurate crystal occurs at higher coverage ($\sim$ 0.100 \AA$^{-2}$). 
\\ \indent
Thin films of parahydrogen (\ph2) molecules are of interest for a number of reasons. Initially, the investigation was driven mainly by the search for the hypothetical superfluid phase at low temperature \cite{gs}, which might be enhanced in reduced dimensionality and/or in the presence of corrugation. There is now a wealth of theoretical results showing that the early prediction of bulk \ph2 superfluidity is incorrect, as it fails to take into account the strong tendency of the system to crystallize at temperatures where Bose condensation and superfluidity ought to take place in a fluid. Crystallization is predicted to occur even in confinement \cite{omiyinka,delmaestro} and/or in disorder \cite{boninsegni05,turnbull,boninsegni15} or reduced dimensions \cite{boninsegni04,boninsegni13}. Indeed, first principle computer simulations based on realistic intermolecular potentials yield evidence of superfluidity at low temperature ($T \sim 1$ K) only in  nanoscale size  clusters of \ph2 comprising fewer than  $N \sim 20$ molecules \cite{sindzingre,fabio,fabio2,mezzacapo08,fabio3}. 
\\ \indent 
Besides its fundamental relevance \ph2 has also practical interest for its fueling potential. In this context, graphene is regarded as one of the most promising materials for the storage of large amounts of \ph2 \cite{tozzini}; this motivates theoretical efforts aimed at understanding how \ph2 layers on graphene. 
Quantum mechanical effects are significant in \ph2, but  considerably less pronounced than in helium; for, despite the light mass of a \ph2 molecule (half that of a helium atom), the much stronger interaction between two \ph2 molecules imparts to the system a markedly more classical behavior. Thus, one might expect the phase diagram of \ph2 on graphene also mimic that on graphite \cite{nho}, and in particular that  commensurate phases predicted for $^4$He should {\em a fortiori} exist in an adsorbed layer of \ph2 as well.
However, the only theoretical study of \ph2 physisorbed on graphene based on Diffusion Monte Carlo (DMC) simulations \cite{soliti} yielded a ground state ($T=0$) monolayer phase diagram featuring only one commensurate phase, namely the C$_{1/3}$, with a first order phase transition to an incommensurate solid layer at higher coverage, close to 0.08 \AA$^{-2}$. 
\\ \indent
We report in this paper results of a first principle numerical study of the phase diagram of a single layer of \ph2 adsorbed on graphene, in the limit of temperature $T\to 0$. We adopted the same microscopic model but a different computational methodology than that of Ref. \onlinecite {soliti}, namely, we used
the continuous-space worm algorithm. This (Monte Carlo)
technique provides accurate estimates of thermodynamic
properties of Bose systems at finite temperature, and has the
distinct advantage of not relying on any a priori input, such
as a trial wave function in the case of DMC. We carried out
simulations down to a temperature T = 1 K, which, as we
show below, is low enough to regard results as representative
of ground-state physics.
\\ \indent 
Our results are in disagreement with the predictions of
Ref. \onlinecite{soliti}. We show that the physical behavior of this system is  actually qualitatively similar to  that predicted theoretically for $^4$He films on graphite and graphene, as well as in  a previous numerical study of the first layer of \ph2 adsorbed on graphite \cite{nho}. Specifically, we find the two expected commensurate crystalline phases, namely the C$_{1/3}$ and the C$_{7/16}$, with a domain wall phase between them, and with a first order phase transition to an incommensurate solid as coverage is increased above that corresponding to the C$_{7/16}$ commensurate crystal. We argue that the failure of Ref. \onlinecite{soliti} to observe the C$_{7/16}$ phase is due to intrinsic limitations and inherent bias of computing methodology adopted therein. Our results indicate that additional commensurate phases may exist, notably one at coverage $\theta=0.0814$ \Am2, also experimentally observed in adsorbed films of D$_2$ on graphite.
\\
Our simulations do not yield evidence of promotion of \ph2 molecules to second layer for coverages at least up to 0.110 \AA$^{-2}$, 
a finding that is inconsistent 
with previous theoretical results for \ph2 on graphite \cite{nho}, as well as with the quoted experimental estimate \cite{boh}, suggesting a second layer promotion coverage of 0.094 \AA$^{-2}$. The possible reasons for this disagreement are discussed below.
\\ \indent
The remainder of this paper is organized as follows: in Sec. \ref{mc} we introduce the model and provide computational details; in   Sec. \ref{sere} we illustrate our results and provide a theoretical interpretation. Finally, we outline our conclusions in Sec. \ref{conc}.
\section{Model and Methodology}\label{mc}
We model our system of interest as in all previous comparable studies, namely Refs. \onlinecite{soliti,happacher}. We consider a collection of $N$ point-like particles (\ph2 molecules) of mass $m$ and spin zero, i.e., in principle obeying Bose statistics; however, because in the physical settings considered here exchanges of molecules are practically absent \cite{boninsegni04}, we regard them as {\em distinguishable} in this study.\\ \indent 
The system is enclosed in a rectangular simulation cell of sides $L_x = 34.08$ \AA, $L_y=36.89$ \AA\ and 
$L_z=40$ \AA\ (the length in the $z$ direction is unimportant), with periodic boundary conditions in all directions. The \ph2 molecules  move in the presence of an external potential arising from a 2D hexagonal lattice comprising $M=480$ carbon (C) atoms, arranged in the $x-y$ plane with $z=0$ and held fixed at lattice positions. The nominal 2D  density (coverage) of \ph2 is $\theta=N/\cal A$, with ${\cal A}=L_x\times L_y$.  
\\ \indent 
The quantum-mechanical Hamiltonian of the system is the following:
\begin{equation}\label{ham}
\hat {\cal H} = -\lambda\sum_{i=1}^N\nabla_i^2 +\sum_{i<j}V(r_{ij})+\sum_{i\sigma}U(|{\bf r}_i-{\bf R}_\sigma|)
\end{equation}
Here, $\lambda\equiv \hbar^2/2m = 12.031$ K\AA$^2$, $V$ is the interaction potential between any two \ph2 molecules, only depending on their relative distance $r_{ij}\equiv |{\bf r}_i-{\bf r}_j|$, whereas $U$ is  the interaction between a \ph2 molecule and a C atom, also depending on their relative distance. The positions ${\bf R}_\sigma$, with $\sigma=1,2,...,M$  are those of the C atoms.
\\ \indent
The results presented here are obtained using the accepted Silvera-Goldman \cite{SG} potential to describe the interaction between two \ph2 molecules. The $U$ term in (\ref{ham}) is modeled by means of a Lennard-Jones potential with parameters $\epsilon=42.8$ K and $\sigma$=2.97 \AA, suggested in Ref. \onlinecite{stan} and adopted in Ref. \onlinecite{soliti}. The use of pair-wise central potentials in (\ref{ham}) is clearly a major simplification from the computational standpoint, justified to the extent that the results can be regarded as still reasonably reliable, at least for specific purposes. Now, the Silvera-Goldman potential has been shown to provide a quantitatively accurate description of the equilibrium solid phase of \ph2 \cite{operetto}. On the other hand, various model interactions have been proposed and used for the \ph2-C part, and the differences among them are often quite substantial \cite{sun}, rendering the choice of any one of them a tricky proposition. 
In this work we made use of the above described Lennard-Jones potential for the purpose of comparing with previous works.
\\ \indent 
We studied the low temperature physical properties of the system described by Eqs. (\ref {ham}) by means of first principle computer simulations based on the Worm
Algorithm in the continuous-space path integral representation \cite{worm,worm2}. Because this well-established computational methodology is thoroughly described elsewhere, we do not review it here. It enables one to compute thermodynamic properties of quantum many-body systems at finite temperature, directly from the microscopic Hamiltonian (including energetic and structural properties of relevance here), in practice with no approximation. Technical details of the simulation are standard, and we refer the interested reader to Ref. \onlinecite{worm2}. We used the standard high-temperature approximation for the many-particle propagator accurate up to order $\tau^4$, and all of the results reported here are extrapolated to the $\tau\to 0$ limit;  in general, we found that a value of the imaginary time step 
$\tau=1/320$ K$^{-1}$ yields estimates that are indistinguishable from the extrapolated ones, within the statistical errors of the calculation. As mentioned above, \ph2 molecules are treated in this work as distinguishable (``Boltzmannons''); thus, in this particular case the Worm Algorithm is very similar to conventional path integral Monte Carlo \cite{jltp}. 
\\ \indent
\section{Results}\label{sere}
\begin{figure}[t]
\centering
\includegraphics[width=\linewidth]{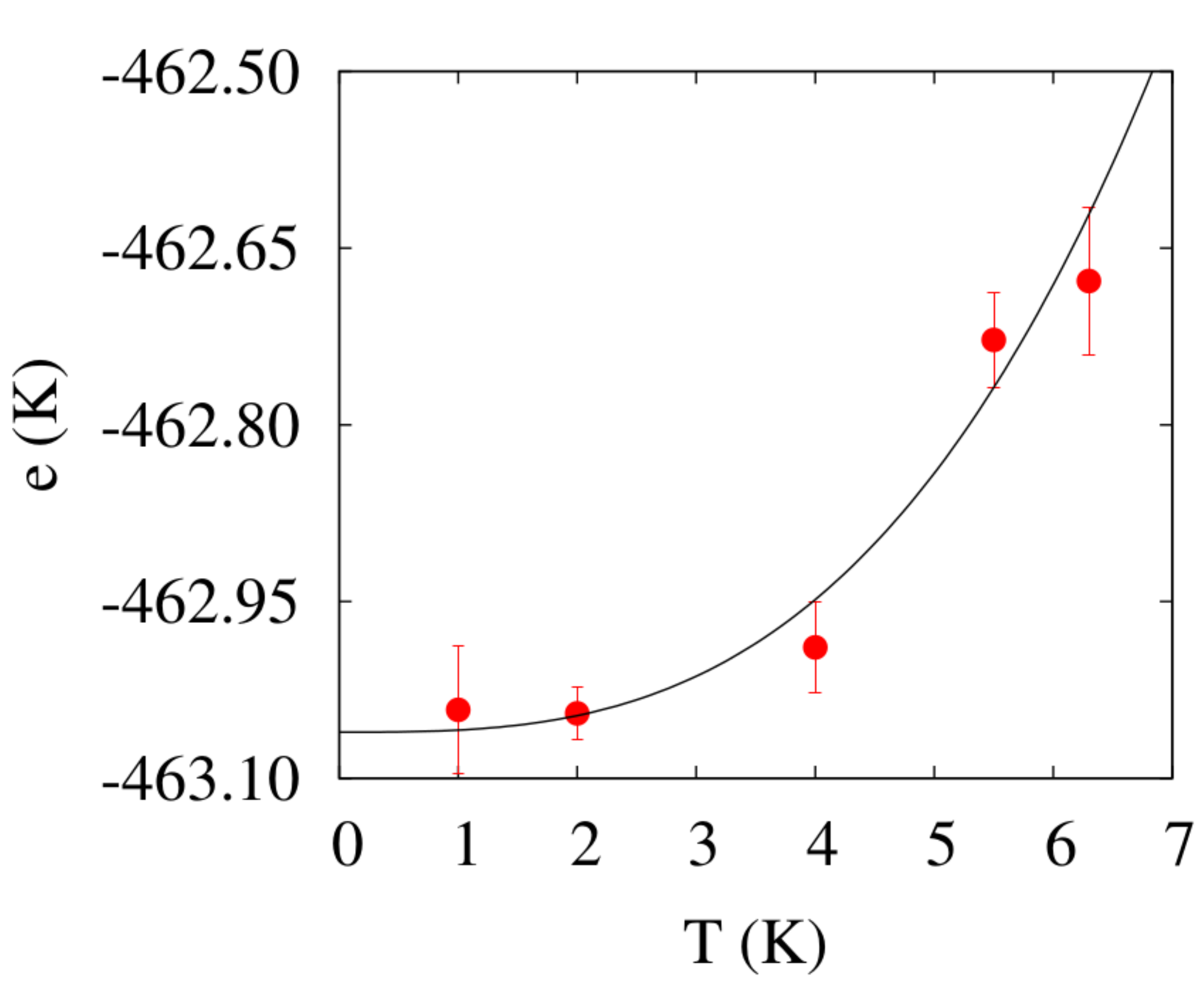}
\caption{{\em Color online.} Energy per \ph2 molecules (in K) as a function of temperature, for a coverage $\theta_\circ=0.0636$ \AA$^{-2}$. Solid line is a fit to the data based on the expression $e(T)=e_\circ+\alpha T^3$. The extrapolated ground state energy per molecule is $-463.06(2)$ K. }
\label{f1}
\end{figure} 

As mentioned in the Introduction, the main objective of this study is the understanding of the $T=0$ phase diagram of a single layer of \ph2 adsorbed on graphene. The expectation is that \ph2 should form crystal structures of different kinds, depending  on the coverage. Insight into this can be gained by computing the ground state energy per \ph2 molecule $e(\theta,T=0)$, as well as by direct observation of the structures that {\em spontaneously} arise at various coverages.
\\ \indent
Because our computational methodology is a finite temperature one,  it becomes necessary to extrapolate to $T=0$ results obtained at finite temperature. Fig. \ref{f1} shows one such extrapolation, specifically at the (1/3) coverage $\theta_\circ=0.0636$ \AA$^{-2}$. 
The finite temperature estimates of the energy per \ph2 molecule $e(T)$ for coverage $\theta_\circ$ are shown, together with a fit to the expression $e(\theta,T)=e_\circ(\theta)+\alpha(\theta) T^3$, which is based on the assumption of 2D phonons being the low-lying excitations of the system. 
The value of $e_\circ$ at this coverage is $-463.06(2)$ K, i.e., a significant downward revision ($\sim 2$ K) with respect to the previous estimate from Ref. \onlinecite{soliti} of $-461.12(1)$ K. Within the statistical errors of the calculation, the value at $T$=4 K is indistinguishable from the extrapolated $T=0$ one. We found this to be consistently the case at other coverages as well, i.e., energy values at $T$=4 K provide a close estimate of ground state energies.
\\ \indent
It should be noted that in the infinite dilution limit (i.e., $\theta\to 0$), the computed single-molecule binding energy is $-432.8(1)$ K (computed at $T=1$ K), about 1 K lower than the ground state value quoted in Ref. \onlinecite{soliti} which is $-431.79(6)$ K. To the best of our determination, the microscopic model utilized here and in Ref. \onlinecite{soliti} are the same, and it is therefore not clear what could be the reason for this small discrepancy. 
Finite temperature techniques have 
time and again 
proven more reliable than ground state ones, when it comes the to ground state of Bose systems \cite{boninsegni15,boninsegni13}. The main advantage that finite temperature methods enjoy is that they are unbiased, i.e., they do not rely on {\em a priori} input, such as a trial wave function, and are not affected by population control bias, unlike DMC \cite{ps,moroni,ps2}. 
\begin{figure}[h]
\centering
\includegraphics[width=\linewidth]{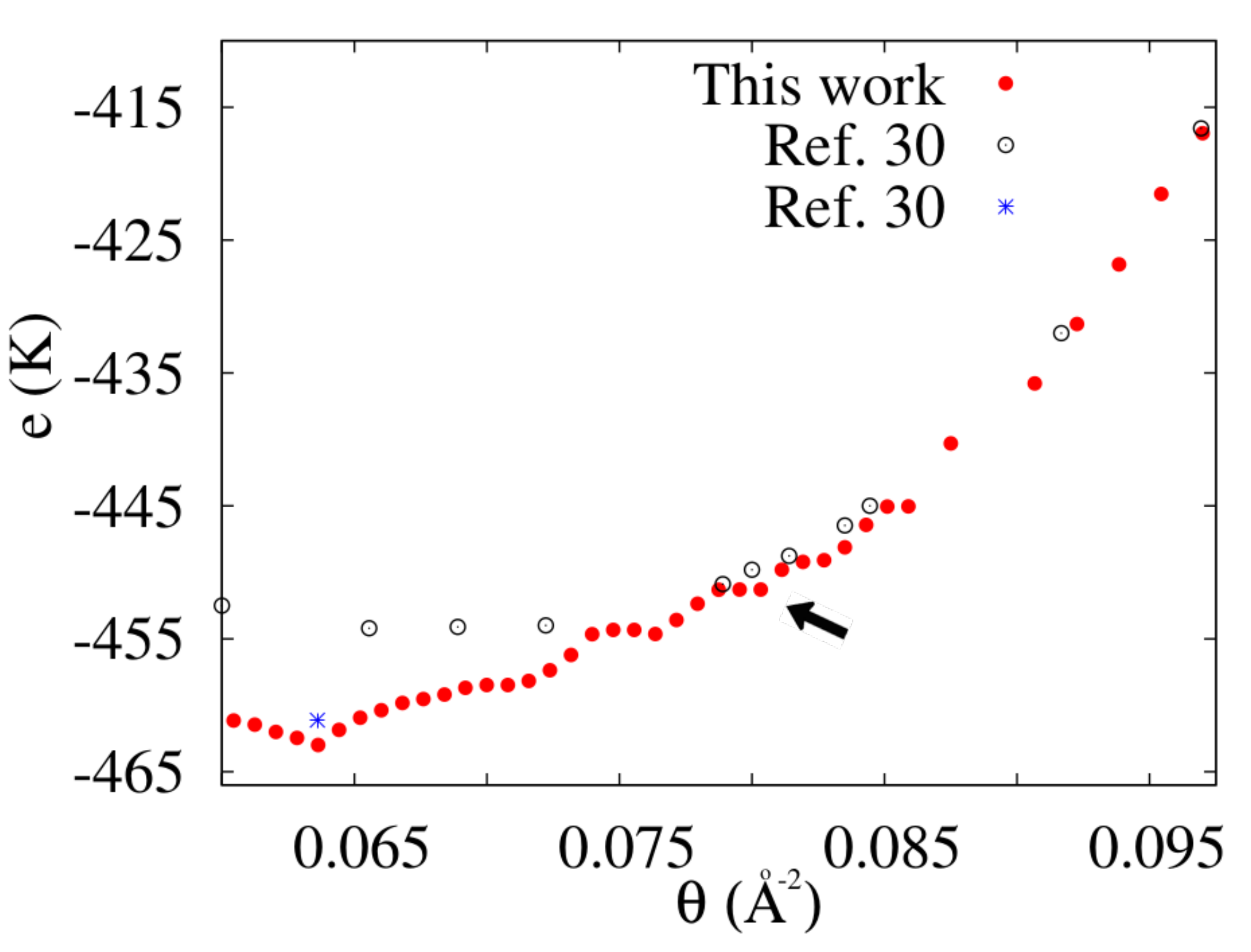}
\caption{{\em Color online.} Energy per \ph2 molecules (in K) as a function of coverage (in \AA$^{-2}$), at temperature $T=4$ K. Open circles (star) are the DMC ground state estimates of Ref. \onlinecite {soliti}, based on a trial wave function assuming incommensurate (C$_{1/3}$ commensurate) crystal order. Statistical errors are smaller than the sizes of the symbols. Arrow points to a possible additional commensurate phase at $\theta=0.0814$ \Am2.}
\label{f2}
\end{figure}
\\ \indent
Fig. \ref{f2} shows the numerical estimates of the ground state energy per particle $e(\theta)$ obtained in this work (filled symbols) at temperature $T$=4 K, as function of coverage. As mentioned above, these energy values remain practically unchanged as the temperature is lowered (we have established this assertion by computing the energy down to a temperature $T=1$ K for a few selected coverages, including the highest one considered here, namely $\theta=0.110$ \Am2). We also show for comparison the ground state estimates from Ref. \onlinecite{soliti},  obtained assuming a crystalline arrangement of \ph2 molecules incommensurate with the underlying substrate (open circles), as well as the formation of a commensurate structure at coverage 1/3 (star).
\begin{table}
\begin{ruledtabular}
\begin{tabular}{ccc}
$\theta$ (\Am2) &Ref. \onlinecite{soliti} ($T=0$) & This work 
($T=4$ K)\\
0.0789 & $-450.88\pm 0.10$ &$-451.28\pm0.04$ \\
0.0814 &$-448.76\pm0.10$  &$-449.58\pm 0.06$ \\
0.0835 &$-446.47\pm0.10$ &$-446.92\pm0.04$\\
\end{tabular}
\end{ruledtabular}
\caption{\label{table1} Energy per \ph2 molecule (in K) as reported in Ref. \onlinecite{soliti} as ground state estimates, and obtained in this work at temperature $T=4$ K, for a 
few coverages near $\theta=0.0835$ \Am2.}\label{tableone}
\end{table}
%
The first observation is that there is quantitative agreement between the energy estimates obtained in this work and the open circles, for $\theta\gtrsim 0.090$ \Am2. For lower coverages, our energy values at $T=4$ K are consistently lower
(a direct comparison with the few values explicitly reported in Ref. \onlinecite{soliti} is offered in Table 
\ref{table1}), the difference becoming rather large (5 K or more) for $\theta\lesssim 0.069$ \Am2 \cite{digi}. This suggests that in this range of \ph2 coverage the physics of the adsorbed layer bears little resemblance to that of an incommensurate crystalline monolayer, or, differently phrased, the corrugation of the substrate plays an important role, not just at {\em exactly} 1/3 coverage. Indeed, the shape of the $e(\theta)$ curve computed here resembles that of the energetics of a film of \ph2 on a graphite substrate obtained in Ref. \onlinecite{nho} using a finite temperature method related to that adopted here, and more generally to what is observed in the presence of commensurate phases \cite{turnbull,crespi}. In particular, there is a minimum at the coverage $\theta_\circ$ corresponding to the C$_{1/3}$ phase, immediately above which the curve starts off with negative curvature. All of this suggests that the evolution of the system as a function of coverage, above $\theta_\circ$, ought to be similar to what observed on graphite, much as for helium films.
\begin{figure}[h]
\centering
\includegraphics[width=\linewidth]{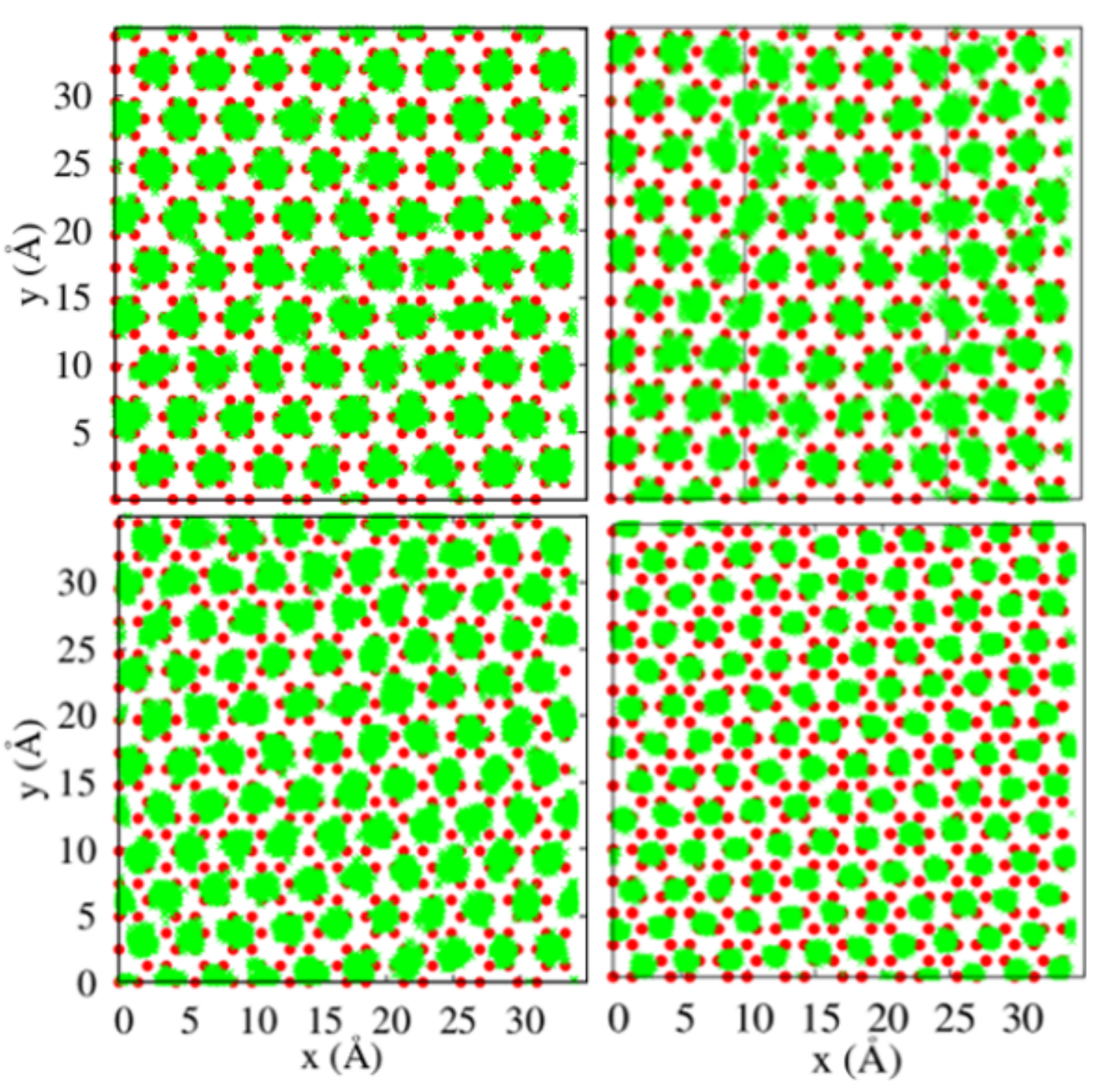}
\caption{{\em Color online.} Configurational snapshots of the system at temperature $T=1$ K, for four different coverages. Shown are the particle world lines, whereas filled circles represent C atoms. {Top Left}: Coverage $\theta_\circ=0.0636$ \Am2 (C$_{1/3}$). {Top Right}: Coverage $\theta=0.0716$ \Am2, vertical lines show the domain walls. {Bottom Left}: Coverage $\theta=0.0835$ \Am2 (C$_{7/16}$). Solid lines show two adjacent unit cells, each containing 7 molecules. {Bottom Right}: Coverage $\theta=0.110$ \Am2.}
\label{f3}
\end{figure}

In order to gain physical insight in the nature of the phases displayed by the system, we have considered in detail four coverages, the same considered in other studies of helium \cite{corboz,happacher} or \ph2 \cite{nho} on either graphite or graphene, in order to detect possible physical differences with respect to the system studied here. As we shall see, everything points to remarkably similar physical behaviors in all these cases.
\\ \indent
Fig. \ref{f3} shows four representative configurational snapshots (particle world lines) of the system at temperature $T=1$ K for the four coverages of interest. By ``representative'' we mean here that such arrangements of molecules, or physically equivalent ones, are consistently observed in the course of sufficiently long simulation runs (it should be mentioned that simulations are started out from high temperature, disordered configurations of molecules)  \cite{notey}.
Top left panel of Fig. \ref{f3} shows the C$_{1/3}$ commensurate structure at coverage $\theta_\circ$. On increasing coverage, analogously to what found for $^4$He on the same substrate \cite{happacher} we observe the appearance of domain walls, as shown in Fig. 2 (top right) for a coverage $\theta=0.0716$ \Am2. One can observe two regions in which the system displays the same structure as in the C$_{1/3}$ phase, but with a relative vertical (i.e., in the $y$ direction) shift of molecules with respect to one another; the two regions are separated by domain walls, featuring increased local \ph2 density. This is also consistent with what observed for \ph2 on graphite \cite{nho}. In the case of $^4$He, as $\theta$ is further increased one observes a proliferation of domain walls, leading to the appearance of a stable, second commensurate phase at coverage $\theta=0.0835$ \Am2; this phase is labeled C$_{7/16}$, and is observed in this work as well (bottom left of Fig. \ref{f3}).
\\ \indent 
The calculation carried out in Ref. \onlinecite{soliti} did not reach a definitive conclusion regarding the existence of the `striped'' phase (e.g., at $\theta=0.0716$ \Am2); on the other hand, the assertion is made therein that the C$_{7/16}$ phase is energetically disfavored with respect to the incommensurate crystalline phase. Our results do not support such a contention, which in our view originates from the failure of the DMC projection algorithm to converge to the true ground state energy for these coverages, as shown in Fig. \ref{f2} and Table \ref{tableone}. 
Here we show that the system does not  directly transition from the C$_{1/3}$ to the incommensurate crystalline phase, as contended in Ref. \onlinecite {soliti}, but rather evolves through at last one other commensurate phase, namely the C$_{7/16}$, as predicted for $^4$He as well \cite{kwon,happacher}.
We did not investigate in detail the possible occurrence of additional commensurate phases in this work, e.g., at coverages $\theta=0.0789$ \Am2 and $\theta=0.0814$ \Am2, which are observed for the heavier isotope D$_2$ adsorbed on graphite \cite{d2}. The sharp feature displayed by the $e(\theta)$ curve at $\theta=0.0814$ \Am2 (shown by the arrow in Fig. \ref{f2}) suggests that this phase is likely present in this system as well, which is consistent with the general observation that the physical behavior of \ph2 on these substrates can be largely understood along classical lines, as already suggested by others \cite{nho}.
\begin{figure}[h]
\centering
\includegraphics[width=\linewidth]{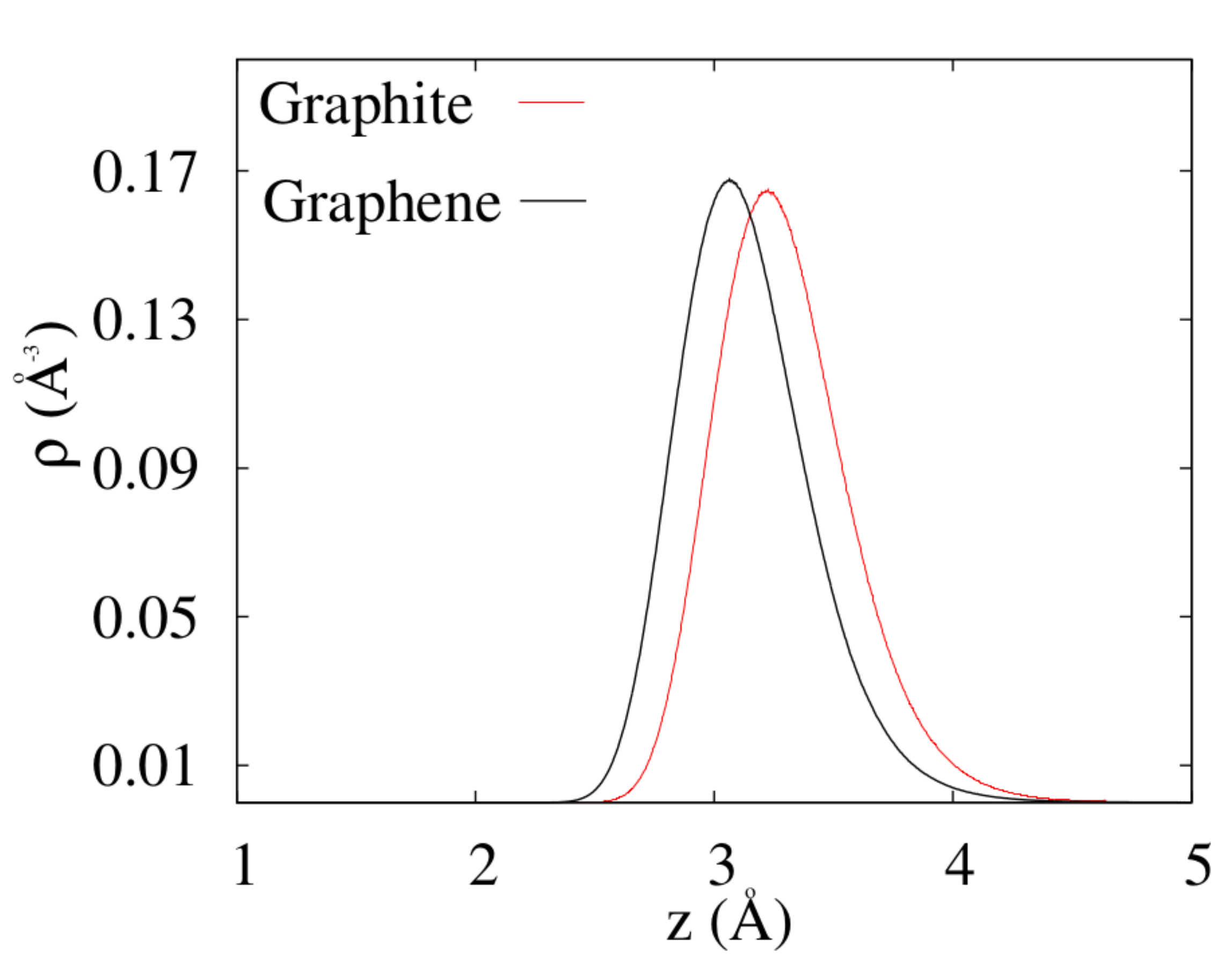}
\caption{{\em Color online.} Density profiles of \ph2 (in \AA$^{-3}$) in the direction perpendicular to the substrate ($z$), in the presence of a graphene and a smooth graphite substrate, at coverage $\theta=0.11$ \Am2. Both profiles are computed at $T=4$ K.}
\label{f4}
\end{figure}
\\ \indent 
The bottom right panel of Fig. \ref {f3} shows a configuration that represents an incommensurate crystalline phase. Such a phase appears to become thermodynamically stable above a coverage of the order of 0.090 \Am2, based on the shape of the $e(\theta)$ curve of Fig. \ref{f2} (the actual determination of the coexistence coverage was not pursued here). A result of our simulation that is in clear quantitative disagreement with existing theoretical predictions \cite{nho} and experimental observation \cite{boh}, has to do with the coverage at which promotion of molecules to the second layer should begin, namely at approximately 0.094 \Am2 for a film adsorbed on a graphite substrate (essentially the same value is predicted on graphene \cite{gb}). It is worth noting that this is a considerably lower coverage  than that at which second layer promotion is observed experimentally \cite{greywall} and predicted theoretically \cite{corboz} for a $^4$He monolayer film adsorbed on graphite (or graphene \cite{happacher}). One would expect a \ph2 monolayer to be more compressible, given the less markedly quantum-mechanical behavior than helium, as well as the much deeper (approximately a factor three) attractive well of the effective potential between a \ph2 molecule and a graphite substrate \cite{crowellb}, with respect to that between the same substrate and a He atom \cite{carlos}; on the other hand, the greater hard core diameter of the interaction between two \ph2 molecules ($\sim 15$\% greater than that between two he atoms) might indeed cause second layer promotion at lower coverages for \ph2.
\\ \indent
In our simulations we observed stable crystalline monolayers of \ph2 of coverage as high as 0.110 \Am2 (bottom right panel of Fig. \ref{f3}), with evidence of promotion to second layer for a coverage $\theta\sim 0.112$ \Am2, virtually indistinguishable from the corresponding value for $^4$He on graphene \cite{happacher}. In order to investigate this issue in greater depth, we repeated the simulations of Ref. \onlinecite{nho}, i.e., for a \ph2 monolayer on a smooth graphite substrate, using the same (Crowell-Brown) potential  used therein, as that was the first numerical studies yielding a second layer promotion coverage in agreement with Ref. \onlinecite{boh}. In this case too, however, we observed a second layer promotion coverage of 0.117(1) \Am2,i.e., our results are in disagreement \cite{add} with those of Ref. \onlinecite{nho}, as well as 20\% higher than current experimental estimates.  The computed \ph2 density profiles in the direction ($z$) perpendicular to the substrate, at a coverage $\theta=0.110$ \Am2 for the cases of both graphene and graphite, are shown in Fig. \ref {f4}. The graphene substrate, in which all the carbon atoms are explicitly included, is less  attractive than the smooth graphite one, witness a $\sim 10\%$ lower molecular binding energy, but in fact molecules are on average closer to the substrate, a fact that likely should be ascribed to the particular forms of the \ph2-substrate interactions adopted here. In neither case, however, do the density profiles yield any indication of possible second layer promotion, as confirmed by visual examination of the configurations. 

\section{Conclusions}\label{conc}
We have carried out a first principle numerical study of the low temperature phase diagram of a \ph2 monolayer adsorbed on a graphene sheet. Our main result is that this phase diagram is very similar to that of \ph2 on graphite, in turn analogous to that of a $^4$He adsorbate on both substrates. In all of these different cases, the strong attractiveness of the substrate renders quantum-mechanical effects quantitatively small in the first layer, whose physical behavior can be understood mostly based on classical considerations. Our results are in disagreement with those of the only previous study of the phase diagram of this system carried out in Ref. \onlinecite{soliti}. However, the conclusions of Ref. \onlinecite{soliti} are based exclusively on energy calculations, and the data reported here clearly point to failure of the projection algorithm utilized therein to approach the true ground state, in a wide range of coverage.
In this study we did not pursue the search for additional commensurate crystalline phases of \ph2 between the C$_{1/3}$ and the C$_{7/16}$, specifically those that are observed on graphite substrate for D$_2$ adsorbates; our energy calculation, however, is consistent with the presence of at least one of them, namely that at $\theta=0.0814$ \Am2.
\\ \indent 
Our simulations show no evidence of promotion of molecules to the second layer for coverages at least up to 0.110 \Am2, significantly greater than the second layer promotion coverage predicted in other theoretical calculations for both graphite and graphene, and higher than what reported experimentally for a graphite substrate. We repeated the calculations carried out in the previous theoretical study of Ref. \onlinecite{nho} for a \ph2 monolayer adsorbed on a smooth graphite substrate, and found a second layer promotion coverage close to 0.117 \Am2, significantly above that obtained in that reference, using the same model and very similar methodology.\\ \indent 
While the disagreement with experiment could merely point to deficiencies in the theoretical model (\ref {ham}), the inconsistencies with other theoretical calculations may have methodological origins. The prediction of a second layer promotion coverage carried out in Ref. \onlinecite{gb} is based on a comparison of energy estimates arrived at in a DMC projection based on different trial wave functions, and could be affected by the same convergence problems pointed out here. On the other hand, the calculation carried out in Ref. \onlinecite{nho}
 is based on essentially the same finite temperature technique utilized in this work, but with a different approximation for the high temperature propagator from that adopted here, and it is worth mentioning that significant discrepancies have been reported in the literature between results for \ph2 obtained using the two approximations, the one utilized here typically yielding more accurate estimates \cite{boninsegni05,mezzacapo08,add}.
 In any case, further investigation is needed in order to resolve this current disagreement between different theoretical calculations, as well as between theory and experiment.

$^\star$ Permanent address: Department of Physics, Memorial University Newfoundland, St. John's, Newfoundland.
\section*{acknowledgments}
This work was supported by the Natural Sciences and Engineering Research Council of Canada. 


\begin{thebibliography}{99}
\bibitem{digi}
The data for the open circles not explicitly reported therein were read off Fig. 2 of Ref. \onlinecite{soliti}.
\bibitem{notey}
For example, in the case of incommensurate structures equivalent configurations are either rigidly translated in space or rotated around an axis perpendicular to the place, going through one of the molecules
\bibitem{add}
Our computed values of the energy per molecule for a \ph2 film adsorbed on a smooth graphite substrate are consistently, significantly lower than those reported in Ref. \onlinecite{nho}. For example, our estimate at $T=4$ K for a coverage $\theta=0.096$ \Am2 is $-448.716(25)$ K versus $-444.78(6)$ K of Ref. \onlinecite{nho}, computed at $T=1$ K.
\end{thebibliography}
\end{document}